\documentclass[aps,pra,twocolumn,superscriptaddress,showpacs]{revtex4}

\usepackage[dvips]{graphicx}

\begin{document}

\title{Critical exponents of the disorder-driven superfluid-insulator
  transition in one-dimensional Bose-Einstein condensates}

\author{J. C. C. Cestari}
\affiliation{Instituto de F\'isica, Universidade Federal do Rio Grande do Sul,
Porto Alegre, RS, Brazil}
\author{A. Foerster}
\affiliation{Instituto de F\'isica, Universidade Federal do Rio Grande do Sul,
Porto Alegre, RS, Brazil}
\author{M. A. Gusm\~ao}
\affiliation{Instituto de F\'isica, Universidade Federal do Rio Grande do Sul,
Porto Alegre, RS, Brazil}
\author{M. Continentino}
\affiliation{Centro Brasileiro de Pesquisas F\'{\i}sicas, Rio de
  Janeiro, RJ, Brazil}

\date{\today}

\begin{abstract}
  We investigate the nature of the superfluid-insulator quantum phase
  transition driven by disorder for non-interacting ultracold atoms on
  one-dimensional lattices. We consider two different cases:
  Anderson-type disorder, with local energies randomly distributed,
  and pseudo-disorder due to a potential incommensurate with the
  lattice, which is usually called the Aubry-Andr\'e model. A scaling
  analysis of numerical data for the superfluid fraction for different
  lattice sizes allows us to determine quantum critical exponents
  characterizing the disorder-driven superfluid-insulator transition.
  We also briefly discuss the effect of interactions close to the
  non-interacting quantum critical point of the Aubry-Andr\'e model.
\end{abstract}


\maketitle

\section{Introduction}
\label{sec:Intro}

A superfluid-insulator transition in a disordered noninteracting
system of bosons at zero temperature is a special type of quantum
phase transition (QPT) \cite{mac}. Instead of the more conventional
competition between different interactions, it is disorder that causes
a drastic change in the nature of the ground state, thus altering the
physical characteristics of the material. A similar type of transition
from a metal to an insulator, usually called the Anderson localization
transition, was first proposed by Anderson \cite{Anderson}, and has
been extensively studied in electronic systems \cite{lee}. In general,
the approach focus on the conductance behavior as the Fermi level
changes in the vicinity of the \emph{mobility edge\/}, which separates
localized and extended one-particle states. In this context, the lower
critical dimension has been determined to be $d_L=2$, which means that
all the states are localized in one dimension for any finite amount of
disorder. Nevertheless, given that the states in a strictly non
disordered system are extended, there is a clear change of regime when
the disordered strength is reduced to zero, which can be characterized
as a QPT. 

In the past decade, enormous progress in the techniques for creating
ultracold atom systems in laboratory settings extended the interest in
the disorder effects and localization to bosonic systems (for recent
reviews, see \cite{fallanir,lagenr,aspectr,palenciar}). For bosons, the
transition is from the \emph{insulator\/} (localized) state to the
superfluid one. It was observed both for laser-speckle disorder
\cite{Billy08} and quasiperiodic optical lattices \cite{Roati08} in
Bose-Einstein condensates of $\,^{87}$Rb and $^{39}$K atoms,
respectively. While speckle disorder comes close to the Anderson-type
\emph{random disorder\/} considered in theoretical approaches,
quasi-periodic lattices present a superposition of the lattice
potential with an incommensurate one and can be viewed as
experimental realizations of pseudo-disorder models like the
Aubry-Andr\'e (AA) model \cite{AA}. The latter also shows superfluid
and localized regimes in one dimension, but the transition between
them occurs at a nonzero critical disorder \cite{AA,ingold}.

Recently, we have investigated numerically the superfluid-insulator
transition in one-dimensional, noninteracting systems of bosons with
these two types of disorder \cite{jardel}. Here we focus on the
scaling properties of the superfluid fraction near the
superfluid-insulator transition, obtaining the relevant critical
exponents.  For random disorder, even though the superfluid phase is
destroyed for arbitrarily weak disorder, we show that the transition
can still be described as a quantum critical phenomenon with
well-defined critical exponents and power-law scaling behavior. The
same happens for the AA model, but the universality classes are
different.

Our starting point is a well-known scaling relation for the
singular part of the superfluid density $\rho_s$ close to a quantum
superfluid-insulator phase transition \cite{Fisher89},
\begin{equation}
\label{eq:rhos}
\rho_s \sim |g|^{\nu(d+z-2)},
\end{equation}
where $g$ measures the distance to the quantum critical point (QCP),
$\nu$ is the correlation length exponent (i.e., the correlation length
diverges as $\xi \sim |g|^{-\nu}$ at the QCP), $d$ is the spatial
dimension, and $z$ is the dynamic critical exponent associated with
the QCP. The superfluid density is directly related to the
\textit{helicity modulus\/} \cite{fisherpai}, and can be viewed as a
measure of the system's response to a phase-twisting field. Thus, it
is natural to interpret the correlation length as a phase-coherence
length. In the insulating phase it should coincide with the
localization length, which measures the spatial extent of the wave
functions. This holds also for disordered metals \cite{lee}.

In a finite system, even at criticality, the correlation length is
limited by the system size $L$, and the finite-size-scaling form of
the superfluid density is
\begin{equation}
\label{eq:rhosfinite}
\rho_s \sim L^{-(d+z-2)} F(L/\xi) = L^{-(d+z-2)} F(L |g|^{\nu}) \,.
\end{equation}
The corresponding relation for the superfluid fraction  ($f_s=L^d
\rho_s$) is
\begin{equation}
  \label{eq:sffscaling}
  f_s \sim L^{-(z-2)} F(L |g|^{\nu}) \,.
\end{equation}
This last equation is suitable to determine the critical exponents
$\nu$ and $z$ from a numerical evaluation of $f_{s}$ for various
lattice sizes, as we do in the following.

\section{Anderson-like disorder}
\label{sec:And}

The usual Hamiltonian describing interacting bosons on a lattice is
known as the Bose-Hubbard Hamiltonian, and is given by
\begin{equation} \label{eq:ham}
  H = \sum_{i}\varepsilon_i n_i + \Omega\sum_{\langle ij \rangle}(
  a^{\dag}_i a_j +a^{\dag}_j a_i ) + \frac{U}{2} \sum_{i} n_i (n_i -
  1) \,,
\end{equation}
where $a^{\dag}_i$ and $a^{}_i$ are the creation and annihilation
operators of a boson at the lattice site $i$, $n_i=a^{\dag}_i a^{}_i$
is the corresponding number operator, each site has a single bound
state of energy $\varepsilon_i$, hopping between sites is restricted
to nearest neighbors, with amplitude $\Omega$, and $U$ is a local
repulsive interaction. In the rest of this paper, energies are
measured in units of the tunneling amplitude $\Omega$. An
Anderson-like disorder \cite{Anderson} is introduced by choosing
random local energies with a uniform distribution in the range
$-\Delta/2 \le \varepsilon_i \le \Delta/2$, so that $\Delta$ is a
measure of the disorder strength.

We carried out a thorough numerical analysis of the above model for
the non-interacting case in one spatial dimension.  Details of this
numerical study are given in Ref.~\cite{jardel}. We recall one of the
main results reported there, namely that the superfluid fraction for a
lattice of size $L$ obeys the relation $f_s^{} =
\exp(-\Delta/\Delta_L)^{4/3}$, where $\Delta_L$ is a characteristic
disorder strength for suppression of superfluidity, which scales with
the lattice size as $ \Delta_L = C\, L^{-3/2} $.  This latter relation
is consistent with the expected value of the critical disorder
strength $\Delta_{c}=0$ for destroying the superfluid phase in the
thermodynamic limit for a one-dimensional system. Furthermore,
defining $g \equiv \Delta - \Delta_{c} = \Delta$, this scaling of
$\Delta_L$ with $L$ is recognized as the finite-size version of the
general relation $\xi \sim |g|^{-\nu}$, immediately yielding the
correlation-length exponent $\nu=2/3$.

\begin{figure}
\begin{center}
\includegraphics[width=0.95\linewidth]{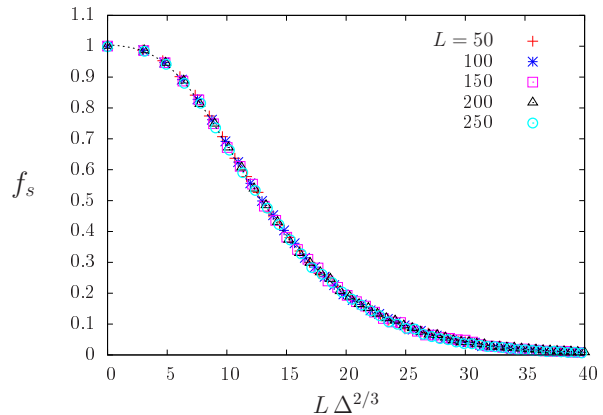}
\end{center}
\caption{(Color online) Finite size scaling of the superfluid fraction
  for Anderson-type disorder. According to Eq.~(\ref{eq:sffscaling}),
  the horizontal-axis variable for collapsed curves is $L |g|^{\nu}$,
  which gives $\nu=2/3$, while the absence of any rescaling of $f_s$
  implies that $z=2$.} \label{fig:sffAnd}
\end{figure}

Equation~(\ref{eq:sffscaling}) implies that $L^{z-2} f_{s}$ is a
universal function of $L\Delta^{\nu}$.  The corresponding plot of our
data for different lattice sizes is shown in Fig.~\ref{fig:sffAnd},
where it is clear that all the data collapse onto a universal
curve. The appearance of $f_s$ alone as the scaling quantity in the
vertical axis means that the dynamic critical exponent is $z=2$. The
scaled variable of the horizontal axis in the collapsed plot confirms
the value $\nu=2/3$ for the correlation-length exponent.  Actually,
for the present problem, we are able to determine explicitly the
scaling function $F(x)$ in Eq.~(\ref{eq:sffscaling}). The above
mentioned expression for $f_s^{}$ implies that
$F(x)=\exp(-x^{2}/C^2)$.  Since $x=L/\xi=L\Delta^{\nu}$, the value
$z=2$ for the dynamic exponent implies a jump of the superfluid
fraction from 0 to 1 at the transition occurring for $\Delta=0$ in the
limit $L \to \infty$. This jump is reminiscent of that of the helicity
modulus in the two-dimensional $XY$ model \cite{harada}.

The value $\nu=2/3$ obtained here for the correlation-length exponent
has not been determined previously, to the best of our knowledge. This
new exponent for the superfluid-insulator transition seems to violate
the inequality $\nu \ge 2/d$, which holds for other disordered systems
\cite{Fisher89}. However, this inequality has been proved only for
interacting systems and for nonzero critical values of the parameter
driving the transition, which is not the case here. On the other hand,
the dynamic exponent $z=2$ implies that the effective dimension of the
quantum phase transition \cite{mac} is $d_{\mathrm{eff}}=d+z=3$.  For
disordered interacting bosons, the Bose-glass-to-superfluid transition
is characterized by the relation $z=d$ \cite{Fisher89}. Therefore,
interacting and noninteracting bosons are in different universality
classes with respect to the localization transition. In the
renormalization group language, interaction is a relevant term close
to the disordered non-interacting fixed point.

\section{Aubry-Andr\'e Model}
\label{sec:AA}

\begin{figure}
\includegraphics[width=0.95\linewidth]{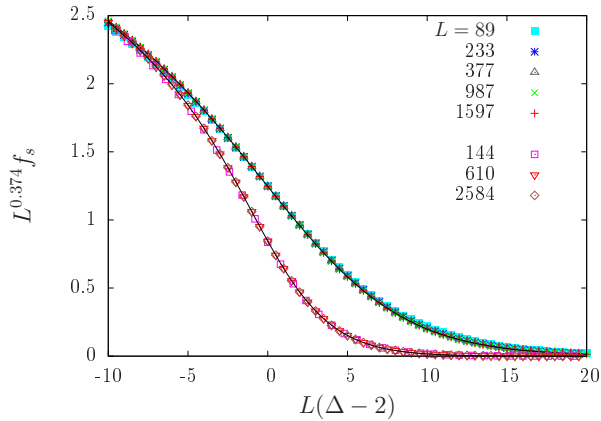}
\caption{(Color online) Finite size scaling of the superfluid fraction
  for the AA model. The data collapse in different curves for even
  and odd numbers of lattice sites (respectively, lower and upper
  curves).} \label{fig:sffAA}
\end{figure}
\begin{figure}
\includegraphics[width=0.95\linewidth,clip]{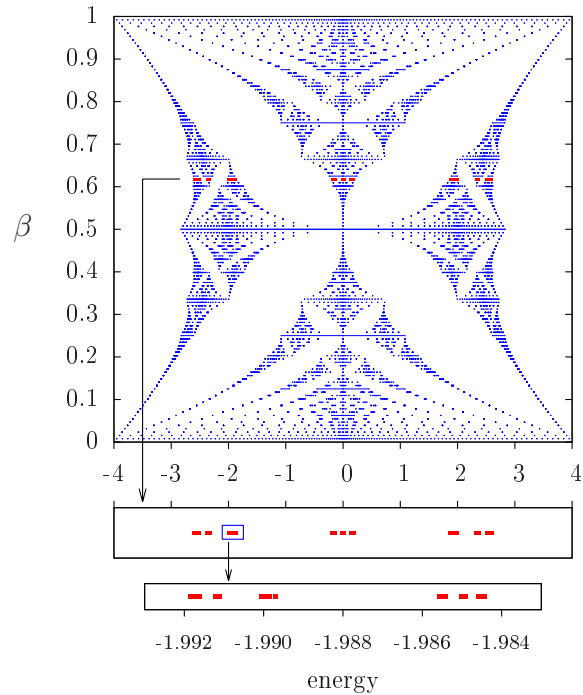}
\caption{(Color online) Spectra of the Harper model, highlighting the
  spectrum corresponding to the AA model (for a rational approximation
  of the golden ratio $\beta=987/610$). Its fractal nature is
  illustrated in the bottom by expanding the small box drawn inside
  the middle panel. We show the spectrum for $\beta -1$,
  which is the same as for $\beta$ according to
  Eq.~(\ref{eq:harper}).} \label{fig:btfly}
\end{figure}

The Aubry-Andr\'e model can also be described by the
Hamiltonian~(\ref{eq:ham}), with $U=0$, except that the distribution
of local energies is not random, but periodic with a period
incommensurate with the lattice spacing. These energies are usually
written as
\begin{equation}
\label{eq:harper}
\varepsilon_i = \Delta \cos(2\pi\beta i),
\end{equation}
where $\beta=(1+\sqrt{5})/2$ is the golden ratio and $i$ assumes
integer values from 1 to $L$.  This is actually a special case of the
Harper model \cite{harper} for electrons in a two-dimensional lattice
in the presence of a perpendicular magnetic field, for which
Eq.~(\ref{eq:harper}) holds for any value of $\beta$, with different
characteristics of the spectrum for rational or irrational
values. Disorder-like effects here are a consequence of the
incommensurability between the ``external potential'' and the
lattice. Aubry and Andr\'e \cite{AA} proved that for this model
localization occurs only when the strength of the potential $\Delta$
is larger than the critical value $\Delta_c = 2$. For finite lattices,
it is convenient to replace $\beta$ with $\beta_n=F_{n+1}/F_n$, the
ratio of two consecutive Fibonacci numbers, whose limit for $n \to
\infty$ is the golden ratio. Then, the lattice size must be chosen as
$L=F_n$ in order to allow for the use of periodic boundary
conditions. For this kind of finite lattices, the critical value
$\Delta_c = 2$ remains a rigorous result \cite{ingold}, since it
corresponds to a duality between the Hamiltonians in position and
momentum space.

It was shown in Ref.~\cite{jardel} that the superfluid fraction
undergoes a very sharp transition around $\Delta=2$ for essentially
all lattice sizes.  This sharpness makes it difficult to directly
extract the correlation-length exponent, as done for random disorder
(Sec.~\ref{sec:And}). Here, we concentrate on a narrow region around
$\Delta_c$, searching for the appropriate scaling variable
proportional to $g \equiv \Delta-\Delta_c = \Delta-2$, and the
appropriate scaling of the superfluid fraction. Our results are shown
in Fig.~\ref{fig:sffAA}. The data collapse onto two universal curves,
for even and odd numbers of lattice sites.  Although the scaling
functions are different for these two cases, the critical exponents
for which the curves collapse are the same. In view of
Eq.~(\ref{eq:sffscaling}) we immediately identify the
correlation-length exponent $\nu=1$ from the $x$-axis scaling variable
in Fig.~\ref{fig:sffAA}, and the dynamic exponent $z=2.374\;$ from the
$y$-axis scaling. It was already known \cite{AA} that $\nu=1$ for this
model. Next we discuss the obtained value of $z$ in the light of
properties of the energy spectrum.

The spectrum of the Harper model has been thoroughly studied in the
past \cite{kohmoto1,Sokoloff,kohmoto2,kohmoto3,kohmoto4}.  For general
rational values of $\beta$ it is multifractal at $\Delta=2$, yielding
the famous \emph{Hofstadter butterfly\/} \cite{Hofstadter}, shown in
Fig.~\ref{fig:btfly}. There we highlight the case that we are studying
here, with $\beta$ being a rational approximant of the golden ratio.
In particular, the figure shows our result obtained from numerical
diagonalization of the Hamiltonian for a lattice of size $L=610$. The
two bottom panels illustrate the fractal nature of this spectrum.

With our replacement of $\beta$ by a ratio of two Fibonacci numbers,
$\beta_n=F_{n+1}/F_n$, the spectrum is equivalent to the one for
$\bar\beta_n = \beta_n - 1 = F_{n-1}/F_n$, which contains $F_n$ bands
and $F_{n-1}$ gaps.  As discussed in detail in
Refs.~\cite{kohmoto2,kohmoto3}, when $F_n = L \to \infty$ the width
$\Delta E_L$ of a given band belonging to the spectrum scales as
$\Delta E_L \sim L^{-\gamma}$, with different regions of the spectrum
associated with different values of $\gamma$ (not to be confused with
the susceptibility critical exponent). In particular, a maximum value
$\gamma_{\mathrm{max}} = 2.374$ corresponds to band-edge states. On
the other hand, the band width is a characteristic energy of the
system and therefore should scale as $\xi^{-z}$, which means that
$\Delta E_L \sim L^{-z}$. Our finding of $z=\gamma_{\mathrm{max}}$ is
in agreement with the relevant state for the zero-temperature
superfluid-insulator transition being the bottom edge of the
lowest-lying band.

\section{Effects of interaction}
\label{sec:ineract}

\begin{figure}
\includegraphics[width=0.95\linewidth]{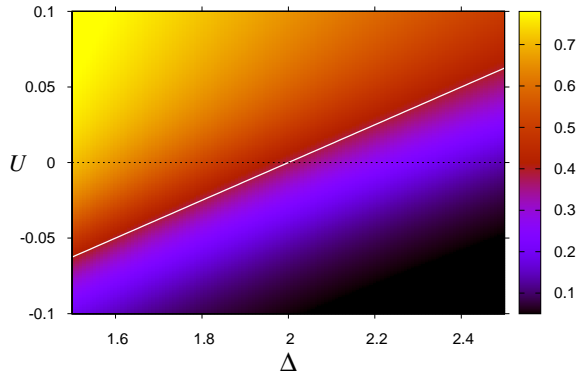}
\caption{(Color online) Zero-temperature phase diagram near the
  localization transition in the presence of a small interaction. The
  color scale indicates values of the superfluid fraction. The
  line $U_c(g)$ in the vicinity of the non-interacting QCP is
  essentially linear (with a slope close to 0.1) implying that it is
  dominated by the analytic part $f(g)$ in
  Eq.~(\ref{eq:Uc_gen}).} \label{fig:UxDel}
\end{figure}

The interaction term in Eq.~(\ref{eq:ham}) can be treated as a
relevant field close to the QCP. The knowledge of the dynamic exponent 
allows us to generalize the scaling relations close to the
QCP for small but finite $U$. The free energy, for example, is given
by
\begin{equation}
\label{eq:freen}
F_s \propto |g|^{\nu(d+z)} P({U}/{|g|^{\nu z}})\,,
\end{equation}
where again we used the fact that $U$ is an energy, and thus scales
with $\xi^{-z}$.  From the above equation, we see that the scaling
contribution to the critical line separating the superfluid and
insulating phases is
\begin{equation}
\label{eq:Uc}
U_c(g) \propto |g|^{\nu z}\,.
\end{equation}
The critical exponents are those associated with the disordered
non-interacting QCP at $U=0$, $\Delta=\Delta_C=2$.  Since the product
$\nu z=2.374$ is large we have to take into account analytic
contributions to the shape of the critical line. This line can be
written in general as
\begin{equation}
  \label{eq:Uc_gen}
U_c = f(g) + a_{\pm} |g|^{\nu z} \,, 
\end{equation}
where $f(g)$ is an analytic function, and $\pm$ refers to the sign of
$g$. Expanding close to the QCP, analytic contributions up to the
second order dominate over the scaling term when $g \rightarrow 0$.
To illustrate this point, in Fig.~\ref{fig:UxDel} we show a phase
diagram close to the noninteracting fixed point at $\Delta_c=2$, for
repulsive ($U>0$) and attractive ($U<0$) interactions. This is a plot
of the superfluid fraction (color scale) as a function of $\Delta$ and
$U$, obtained by diagonalizing the Hamiltonian~(\ref{eq:ham}) for
$N=8$ interacting bosons on a lattice of $L=8$ sites.  We can see that
a straight line [i.e., $f(g) \sim g$] is a very good approximation to
the boundary between the superfluid and localized regions. Even though
the transition is smoothed out by the small lattice size, it is worth
mentioning that the value of $f_s$ at the critical point
$(U=0,\Delta=2)$ is compatible with the lower curve of
Fig.~\ref{fig:sffAA} for $L=8$.

\section{Conclusions}
\label{sec:concl}

We studied the superfluid-insulator transition for bosons on a
one-dimensional lattice, both with random disorder and the pseudo
disorder described by the Aubry-Andr\'e model, the two prototype
models employed in the investigation of localization for ultracold
atoms in optical lattices. Using a finite-size-scaling analysis of the
superfluid fraction, we obtained the critical exponents characterizing
this transition.  The superfluid fraction yields the
correlation-length exponent $\nu$ and the dynamic critical exponent
$z$. For random disorder we found $\nu=2/3$ and $z=2$, while for the
AA model the results are $\nu=1$ and $z=2.374$. The other critical
exponents can be obtained from the quantum hyperscaling relations
\cite{mac} $2-\alpha=\nu(d+z)$ and $2 \beta=\nu(d+z-2+\eta)$. These
two models fall into different universality classes, which is not
surprising since the critical disorder strength for the
superfluid-insulator transition is zero for Anderson-like disorder and
nonzero for the AA model, which also exhibits a multifractal energy
spectrum at the QCP.

It is interesting to observe that the scaling form of the free energy
for nonzero temperature $T$ can be used to determine the
thermodynamic behavior close to the superfluid-insulating QCP. From
it, a general dependence of the specific heat with temperature is
obtained \cite{Fisher89}, with the form $C \sim T^{d/z}$.

Our brief discussion of interaction effects in the AA model shows that
the critical point moves to stronger disorder for repulsive
interaction, and to weaker disorder in the attractive case. Regions of
the phase diagram that correspond to localized and superfluid regimes
in the thermodynamic limit are separated by a line that is
approximately linear, reflecting the dominance of nonsingular
contributions.

\subsection*{Acknowledgments}

This work was supported in part by CNPq, Conselho Nacional de
Desenvolvimento Cient\'ifico e Tecnol\'ogico (Brazil). M.A.C. also
thanks FAPERJ, Funda\c{c}\~ao de Amparo \`a Pesquisa do Estado do Rio
de Janeiro, for partial financial support.

\end{document}